\documentclass[twocolumn,10pt]{tsfp}
\usepackage{flushend}
\usepackage{graphicx}
\usepackage[authoryear,round]{natbib}
\usepackage{fancyhdr}
\usepackage{csquotes}
\usepackage{amsmath}
\usepackage{bm}      
\usepackage{xcolor}
\usepackage{tikz,pgfplots,siunitx}
\usepackage{pgfplotstable}

\title{Uncovering Turbulent Dynamics in Stenotic Flows from 4D-flow MRI Measurements via Resolvent Analysis and Data Assimilation}

\author{A. Villi{é}
    \affiliation{
	Laboratory for Flow Instability and Dynamics\\
	Technische Universit{ä}t Berlin\\
	10623, Berlin, Germany\\
    Contact: a.villie@tu-berlin.de
    }	
}
\author{H. Dillinger, S. Schmitter
        \affiliation{
	Physikalisch-Technische Bundesanstalt\\ 
    Braunschweig and Berlin\\
    10587 Berlin, Germany
    }	
}
\author{S. Demange, K. Oberleithner
    \affiliation{
	Laboratory for Flow Instability and Dynamics\\
	Technische Universit{ä}t Berlin\\
	10623, Berlin, Germany\\
    }	
}

\newcommand{\mdot}{\Dot{m}}

\def\tr[#1]{\textcolor{red}{#1}}
\def\tg[#1]{\textcolor{green}{#1}}
\def\tb[#1]{\textcolor{blue}{#1}}

\newcommand{\bff}{{\boldsymbol{f}}}

\newcommand{\bfq}{{\boldsymbol{q}}}

\newcommand{\bfu}{{\boldsymbol{u}}}

\newcommand{\bfx}{{\boldsymbol{x}}}

\newcommand{\bfy}{{\boldsymbol{y}}}

\newcommand{\bau}{\overline{u}}
\newcommand{\babfu}{\overline{\bfu}}
\newcommand{\babfq}{\overline{\bfq}}

\newcommand{\bap}{\overline{p}}

\def\calL{\mathcal{L}}

\newcommand{\hbfq}{\boldsymbol{\hat{q}}}

\newcommand{\hbff}{\boldsymbol{\hat{f}}}
\newcommand{\hbfu}{\boldsymbol{\hat{u}}}

\makeatletter
\newcommand{\eqautoref}[1]{%
  \begingroup
  \def\equationautorefname~##1\null{%
    \ifnum\count@=1
      Eq.~(##1)%
    \else
      Eqs.~(##1)%
    \fi
    \null
  }%
  \eqautoref@parse#1,\relax
  \endgroup
}
\def\eqautoref@parse#1,#2\relax{%
  \count@=0
  \eqautoref@item{#1}%
  \ifx\\#2\\%
  \else
    ,\eqautoref@parse#2\relax
  \fi
}
\def\eqautoref@item#1{%
  \advance\count@ by 1
  \autoref{#1}%
}
\makeatother

\begin{document}

\maketitle   
\thispagestyle{fancy}

\fontsize{9}{11}\selectfont

\setlength{\belowdisplayskip}{10pt} \setlength{\belowdisplayshortskip}{10pt} \setlength{\abovedisplayskip}{10pt} \setlength{\abovedisplayshortskip}{10pt}
\section*{ABSTRACT}
This study presents a hybrid experimental and computational framework that couples \textit{in vitro} 4D phase-contrast magnetic resonance imaging (4D-flow MRI) measurements with data assimilation and linear modeling to characterize the flow linear amplification mechanisms. 
We manufacture an idealized stenosis phantom with a cosine-shaped contraction and acquire three-dimensional (3D) mean velocity measurements at Reynolds number $3960$ using 4D-flow MRI. 
To overcome the inherent displacement artifact, we perform data assimilation via a two-step optimization strategy using physics-informed neural network (PINN).
This approach first corrects measurement artifacts 
before extracting the unknown mean pressure and eddy viscosity fields. 
The RANS-compatible mean flow then serves as the base state for global linear stability analysis (LSA) and resolvent analysis.
The global LSA reveals stationary eigenmodes located in the recirculation bubble that exhibit a positive growth rate for azimuthal wavenumbers $m=2$ and $m=3$.
The forced dynamics of this eigenmode dominates the low-frequency dynamics. 
Resolvent analysis identifies a broadband pseudo-resonance associated with the convective instability of the separated shear-layer, with maximal amplification for $m=0$.
This methodology demonstrates how integrating sparse experimental MRI data with physics-based modeling enables the identification of mean fields and coherent structures. By leveraging the capabilities of 4D-flow MRI to non-invasively measure 3D velocity fields without requiring physical or optical access, this approach is a first step in the application of linear analysis to cardiovascular flows.

\section*{INTRODUCTION}
The dynamics of blood flow past arterial stenoses centrally influence the development and progression of cardiovascular diseases. Atherosclerosis induces a localized narrowing, or stenosis, which can cause non-recoverable pressure loss and flow choking. The downstream disturbed flow regions exhibit low-magnitude and oscillatory shear stresses, which correlate with arterial wall thickening and endothelial dysfunction \citep{ku1997blood}. 
Experimental studies show that vortex shedding from the constriction throat drives the onset of unsteadiness and transition to turbulence \cite{vetel2008asymmetry}. 
\citet{sherwin2005three} carried out a global stability analysis and found a stationary eigenmode that becomes unstable above the critical Reynolds number $Re_c = 722$, that they interpret as a Coanda-type instability.
\citet{blackburn2008convective} later performed transient growth analyses of laminar base flows to characterized the onset of unsteadiness in stenotic flows. They have demonstrated that the primary route to transition is driven by convective instability at even smaller Reynolds numbers, identifying sinuous perturbations in the separated shear-layer as the most energetic structures.
However, these stability studies did not investigate dynamics beyond a Reynolds number of $1000$. 
Given that peak systolic Reynolds numbers in the aorta can reach $4000$ \citep{ku1997blood}, characterizing the persistent flow structures driven by continuous nonlinear forcing under turbulent conditions remains an unexplored yet essential step for deepening our understanding of atherosclerosis.

4D-flow Magnetic Resonance Imaging (4D-flow MRI) is a non-invasive measurement method that can help characterize complex hemodynamics in stenotic flows \citep{markl20124d}. It allows to resolve 3D velocity fields in complex geometries, but 
suffers from inherent resolution constraints that does not allow to capture the turbulent scales. 
Furthermore, the displacement artifact, or misregistration error, constitutes primary source of error \citep{Busch2013}. Because an encoding delay separates the acquisition of each velocity component, the moving fluid shifts the measured velocities spatially, corrupting the accuracy of the velocity field \citep{dillinger2020limitations}. For steady laminar flows, a correction method is available \citep{Thunberg2002} which, however, is not applicable to unsteady or turbulent flows. 

Physics-Informed Neural Networks (PINNs) provide a flexible physics-augmented framework for data assimilation \citep{raissi2019physics}.
It approximates data with a continuous function, the neural network, that additionally approximates a solution to a given set of PDEs. The training includes physical information into the network by optimizing composite loss functions which, in addition to the typical data loss term, contain an additional physical loss term composed of the PDEs residuals.
Because the PINN evaluates flow quantities continuously across the spatial domain, they naturally handle sparse and low-resolution experimental data. 
PINN has demonstrated significant potential for enhancing fluid flow measurements through data assimilation, successfully assimilating turbulent mean flow data \citep{von_saldern_mean_2022} and specifically improving stenotic mean flow measurements \citep{arzani_uncovering_2021}. 
In \citet{villie2025physics}, a data assimilation framework using PINN was developed to correct the displacement artifact and to assimilate a high-quality mean field from noisy 4D-flow MRI data.

Correcting the MRI measurement artifacts leads to a high-quality 3D mean field that can serve as a base state for linear modeling. It enables to extract the dominant coherent structures without requiring highly resolved time-series data, and potentially grants access to amplification mechanisms and driving instabilities. 
Global linear stability analysis (LSA) linearizes the Navier-Stokes equations around the mean flow to characterize the intrinsic modal instabilities of the unforced system \citep{schmid2001stability}. 
However, in many separated flows, such as those found in severe stenoses, transition to turbulence arises from convective instabilities and non-modal transient growth rather than unstable global modes \citep{blackburn2008convective, griffith2008steady}. 
To capture these mechanisms, resolvent analysis (RA) formulates an input-output framework that models the nonlinear fluctuation terms as an external harmonic forcing acting on the mean flow \citep{mckeon2010critical}. 
Despite its success in various canonical shear flows, resolvent analysis has not yet been applied to 4D-flow MRI data, nor has it been used in the broader context of cardiovascular flows.

This study leverages 4D-flow MRI data to investigate the stability properties and amplification mechanisms in a stenotic flow at a Reynolds number of $3960$ with a $75\%$ area reduction.
The paper first presents the experimental setup, then introduces the two-step data assimilation framework that corrects the artifact and assimilates the mean flow. Finally, the results of mean flow stability analysis are discussed.

\section*{4D-FLOW MRI MEASUREMENTS}\label{sec:4dflow}
\subsection*{Experimental Setup}
For experimental data, an axisymmetric cosine-shaped stenosis phantom with an areal reduction of 75\% (severe stenosis) was manufactured by 3D stereolithography printing. 
The fluid used was water with molecular viscosity $\nu_m~=~
1.0~\times~10^{-6}~$m$^2$/s. 
A constant flow rate of $\mdot~=~94\times10^{-6}$~m$^3$/s was set at the inlet by a gear pump (CardioFlow5000 MR, Shelly Medical Imaging Technologies). The resulting Reynolds number based on the unobstructed diameter was $Re =3960$.
In the following, we normalize the results by the obstructed diameter $d_t = 0.015$~m and bulk velocity at the stenosis throat $U_t = 0.53$~m/s. 

MRI data is acquired at $3$T (Cima.X, Siemens Healthineers, Erlangen, Germany) using a knee coil and a vendor-provided 4D-flow sequence for velocity measurement. 
The symmetry stenosis axis is aligned with the MRI longitudinal ($z$) axis, so that each measurement slice is perpendicular to the axial flow component.
The data covers a field-of-view of $256$x$144$x$1~\mathrm{mm}^3$ with a resolution of $1$~mm isotropic.
The acquisition incorporates $24$ averages for a total time of $76$~min. Furthermore, the velocity encoding strength is set to $75$cm/s and the echo time was $3.08$~ms.

\subsection*{Post-Processing}

The fully sampled 4D-flow MRI data are inverse Fourier transformed, coil combined by singular value decomposition. 
Background phase correction is performed by subtracting the phase of a scan with one average and the pump being turned off. 

\begin{figure}%
    \centering
    \def\svgwidth{1\linewidth}
    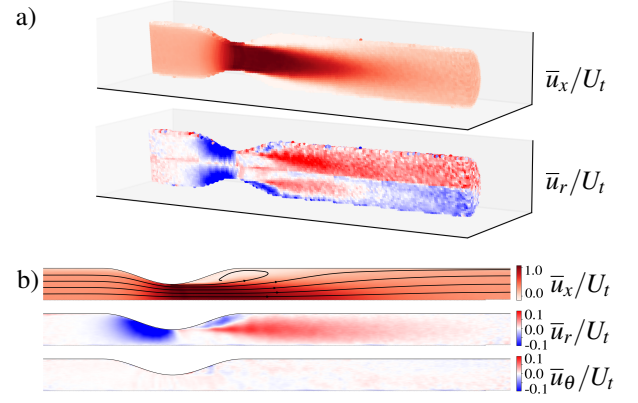
    \caption{Three-dimensional mean flow after segmentation, time-averaging, and transformation to a cylindrical reference frame (a). The azimuthal average of this 3D field yields the 2D mean velocity field (b). The streamlines are plotted on top of the axial component.}
    \label{fig:Exp_flow}
\end{figure}

The fluid volume is segmented using a threshold on the velocity magnitude data. 
The dataset is realigned to coincide with the MRI longitudinal axis. The misalignment angles are calculated by minimizing the spatial standard deviation of the segmented fluid voxels in the transverse directions, and the spatial coordinates and velocity vectors are rotated.
The aligned Cartesian velocity components $(\bau_x, \bau_y, \bau_z)$ are transformed into a cylindrical frame of reference $(\bau_x, \bau_r, \bau_\theta)$, the new axial direction $x$ being aligned with the longitudinal MRI axis. The axial and radial components of this 3D field are shown in figure~\ref{fig:Exp_flow}(a). 

We expect statistical homogeneity in azimuthal direction $\theta$ since the flow has no preferred sense of rotation. To obtain a 2D velocity field in the meridional plane, the rotated MRI points were linearly interpolated onto a structured cylindrical mesh. The final mean flow $\babfu(x, r) = (\bau_x, \bau_r, \bau_\theta)$ is obtained by averaging the interpolated 3D field over the azimuthal direction. Points closest to the known stenosis wall are discarded to exclude any possible voxel that includes solid parts of the phantom.
The 2D mean field is shown in figure~\ref{fig:Exp_flow}(b). We obtain an azimuthal velocity field close to zero, as expected, and set $\bau_\theta = 0$ for the rest of the analysis. The streamlines depict a flow separation at the stenosis throat, forming an axisymmetric recirculation bubble. However, the lower signal-to-noise ratio of the radial component generates a noticeably noisy field. Furthermore, the recirculation streamline appears to spiral rather than form closed loops.

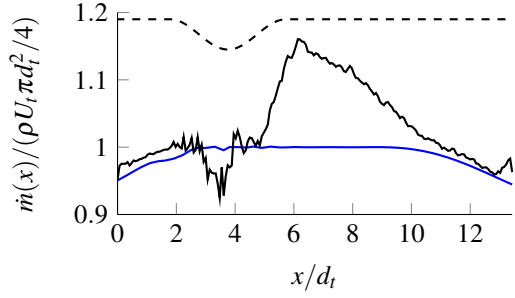
\begin{figure}%
    \centering
    \pgfplotstableread[col sep = comma]{flow_rates_PINN.csv}\mytablepinn
\pgfplotstableread[col sep = comma]{flow_rates_data.csv}\mytabled
\definecolor{myorange}{RGB}{218, 124, 48}

\begin{tikzpicture}
    \begin{axis}[
        width=0.4\textwidth,
        height=0.25\textwidth,
        xmin=0, xmax=13.427,
        ymin=0.9, ymax=1.2,
        xlabel={$x/d_t$},
        ylabel={$\dot{m}(x)/(\rho U_t\pi d_t^2/4)$},
        legend style={at={(0.5,-0.3)}, anchor=north, legend columns=2},
        axis lines*=left,
        grid=none
    ]

    \addplot[blue, solid, thick] table[x expr=\thisrow{x}*13.427, y=FR] {\mytablepinn};
    \label{line:FR_PINN}

    \addplot[black, solid, thick] table[x expr=\thisrow{x}*13.427, y=FR_data] {\mytabled};\label{line:FR_data}
    
    \addplot[black, dashed, thick] table[x expr=\thisrow{x}*13.427, y=upB] {\mytabled};\label{line:upB}

    \end{axis}
\end{tikzpicture}
    \caption{Mean flow rate of the 4D-flow MRI measurements (\ref{line:FR_data}) and mean flow rate after the PINN artifact correction (\ref{line:FR_PINN}) normalized by the theoretical flow rate. The stenosis wall is plotted as a reference (\ref{line:upB}).}
    \label{fig:Flow_rate}
\end{figure}

Before performing mean flow stability analysis, data assimilation is required to correct the displacement artifact. 
This artifact is a spatial offset of the measured velocities caused by an encoding delay between the acquisition of each velocity component.
To illustrate this shift, figure~\ref{fig:Flow_rate} plots the mean flow rate, defined as $\mdot(x)~=~\int^{R(x)}_{0} 2 \pi r \babfu_x(x,r)\text{dr}$, where $R(x)$ denotes the stenosis radius at location $x$. The displacement artifact manifests as a distinct drop and subsequent peak in the measured mean flow rate (black line, \ref{line:FR_data}). The flow acceleration at the stenosis entrance locally underestimates the velocity, whereas the downstream deceleration of the confined jet overestimates the axial velocity.

\section*{DATA ASSIMILATION WITH PINN}\label{sec:DA}
The physics-informed neural network (PINN), maps spatial coordinates $\bfx$ to flow variables $\bar{\bfy}(\bfx)$. The network parameters are optimized by minimizing composite loss functions, including data ($\calL_{\text{data}}$), physics ($\calL_{\text{PDE}}$) and boundary condition ($\calL_{\text{BC}}$) residuals. 

The two objectives of the PINN assimilation are first to correct the displacement artifact, and second to assimilate unknown quantities such as the mean pressure and eddy viscosity by assimilating the Reynolds-averaged Navier-Stokes equations (RANS).
Performing both artifact correction and RANS assimilation in one step from biased experimental data is very challenging. This process easily falls into unsatisfactory local minima, resulting in non-physical fields. The optimization problem is divided into two sequential steps that are easier to solve, corresponding to the two objectives, following closely \citet{villie2025physics}.

\subsection*{Step 1: Artifact Correction}
First, we assimilate a divergence-free mean velocity field to correct the displacement artifact.

\begin{figure}
\centering
\includegraphics[width=1\linewidth]{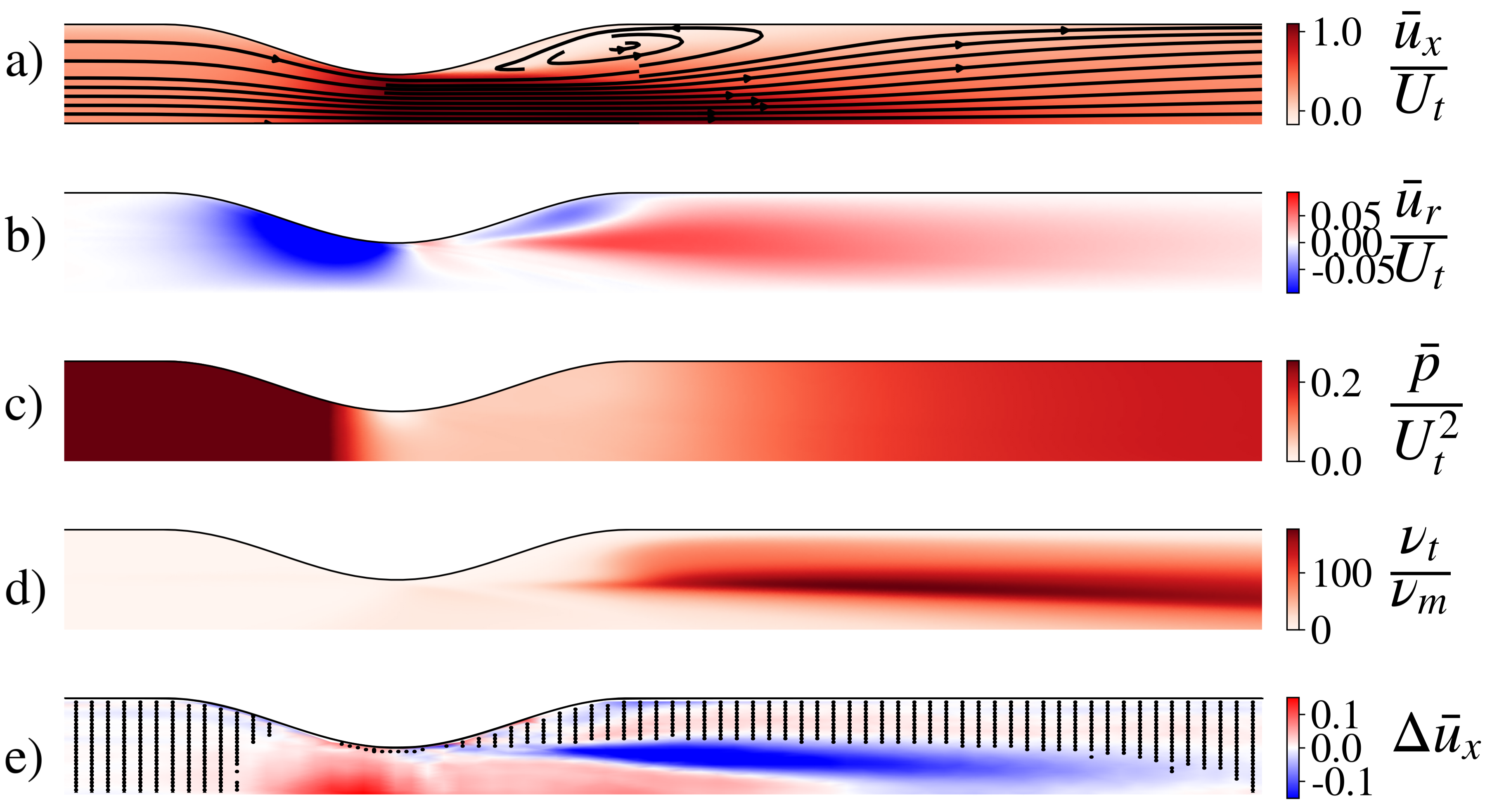}
\caption{PINN-assimilated mean axial and radial velocities (a,b), mean pressure (c) and eddy viscosity (d). The velocity streamlines are plotted over the axial velocity field. Panel (e) illustrates the PINN correction on the axial velocity measurements, along with the training data points for the first optimization step (black dots). }
\label{fig:PINN_mean_flow}
\end{figure}

As the artifact mostly affects location with strong flow velocity, we discard all training points where the axial velocity is above a selected threshold, i.e. $\bau_x > 0.65 \max(\bau_x)$, thus omitting the biased region. The location of the $2144$ training points for the first optimization is shown in figure~\ref{fig:PINN_mean_flow}(e).

The physics loss incorporates two equations on two distinct domains. The continuity equation is optimised over the $2000$ collocation points $\bfx_c$, uniformly distributed in the domain. The flow rate is computed at $20$ axial positions $\bfx_{\mdot}$ equally distributed in the range $[2.5 \, d_t, 11 \, d_t]$, and enforced to remain constant equal to the theoretical flow rate. The physics loss term is:
\begin{equation}
    \calL_{\text{PDE}} =\, \left\| \boldsymbol{\nabla}\cdot \babfu(\bfx_c) \right\|
    + \left\| \mdot(\bfx_{\mdot}) - U_t \pi d_t^2/4\right\| \label{eq:pde-loss}
\end{equation}
The magnitude of the composite losses is computed using the squared L$^2$-norm.
The boundary loss term includes no-slip at the wall for the mean velocity field and axisymmetry at the axis.

We employ a fully connected neural network consisting of $4$ layers with $128$ neurons each. 
The PINN is implemented in Python using TensorFlow, and training is performed with a full-batch limited-memory Broyden–Fletcher–Goldfarb–Shanno (L-BFGS) optimizer.

The PINN-assimilated 2D velocity field is presented in figure~\ref{fig:PINN_mean_flow}(a,b), which the intrinsic denoising behavior of the network automatically smooths. The assimilation correctly closes the streamlines within the recirculation bubble, improving the raw measurements where these streamlines unphysically intersect the wall.
To better visualize the changes in the flow field, we define the absolute data correction $\Delta \babfu$ as the difference between the PINN output and the measured velocities.  Figure~\ref{fig:PINN_mean_flow} shows the axial component of this correction. The corrections are mainly localized in the artifact region with no training data point. Red zones ($\Delta \bau_x < 0$) indicate a reduction in axial velocity, whereas blue zones ($\Delta \bau_x > 0$) reveal flow acceleration. 
This correction emerges from the constraint of a constant mass flow rate~\eqref{eq:pde-loss}. Consequently, the first optimization successfully restores a constant flow rate across this corrected region, as displayed in figure~\ref{fig:Flow_rate} (blue line \ref{line:FR_PINN}). 

\subsection*{Step 2: Mean Pressure and Reynolds Stress Assimilation}
In a second step, we use the PINN-corrected velocity field from the first optimization as training data to assimilate the mean pressure $\bap$ and the eddy viscosity $\nu_t$. 
As a result, the data loss now fixes the velocity field to the first optimization output.

The physics loss term evaluates the residuals of the RANS momentum equations for incompressible flows in cylindrical coordinates over the uniformly distributed collocation points $\bfx_c$:
\begin{align}
    [\calL_{mx}, \calL_{mr}]^T &= \left\|(\babfu \cdot \boldsymbol{\nabla})\babfu + \boldsymbol{\nabla} \bap - \frac{1}{Re} \boldsymbol{\Delta} \babfu + \boldsymbol{\nabla} \cdot \overline{\bfu'\bfu'} \right\|,
\end{align}
where $\overline{\bfu'\bfu'}$ is the Reynolds stress tensor (RST). Directly assimilating the RST components would lead to an underconstrained problem, as the number of unknown variables exceeds the number of governing equations.
To avoid this, we model the deviatoric component of the RST using an eddy viscosity $\nu_t$ that accounts for turbulent dissipation \citep{vonsaldernRoleEddyViscosity2024}, while absorbing the isotropic stress into the modified pressure term $\bap$. The two momentum equations thereby enable the assimilation of [$\bap, \sqrt{\nu_t}$]. Setting the network output to $\sqrt{\nu_t}$ naturally enforces the positivity of $\nu_t$ in the equations. 
%
We update the boundary loss to retain the previous velocity conditions while adding $|\nu_t| = 0$ at the wall and $|\partial\nu_t/\partial r| = |\partial \bap/\partial r| = 0$ at the axis.

The PINN-assimilated modified pressure $\bap$ and $\nu_t$ are presented in figure~\ref{fig:PINN_mean_flow}(c,d).
The mean pressure is minimal at the stenosis throat and slowly recovers downstream of the constriction. It is qualitatively similar to the validation with RANS simulation done in \cite{villie2025physics} for a similar flow. The eddy viscosity field is dominant in the shear-layer.

This enhanced mean flow is now used as a base state in linear modeling to identify linearly amplifying coherent structures and the underlying physical mechanisms.
\section*{LINEAR MODELING} \label{sec:linModel}
This section presents the linear analysis of the assimilated mean flow to model the coherent fluctuations.

The state vector $\bfq(x, r) = [\bfu(x, r), p(x, r)]^T$ is decomposed into mean and fluctuating components, $\bfq = \babfq + \bfq'$. Substituting this Reynolds decomposition into the incompressible Navier-Stokes equations and subtracting the temporal mean yields the perturbation equations:
\begin{eqnarray}
\label{eq:NS_perturbation1}
& \boldsymbol{\nabla} \cdot \bfu' = 0,\\
\label{eq:NS_perturbation2}
& \frac{\partial \bfu}{\partial t}+(\babfu \cdot \boldsymbol{\nabla}) \bfu' + (\bfu' \cdot \boldsymbol{\nabla}) \babfu + \boldsymbol{\nabla} p'-\frac{1}{{Re}}\,(1+\frac{\nu_t}{\nu_m})\,\boldsymbol{\nabla}^2\bfu' = \bff' \notag\\
& \hfill 
\end{eqnarray}
where $\nu_t$ is the eddy viscosity from the PINN and $\bff'$ is the turbulent forcing that includes the nonlinear terms. 

\subsection*{Linear Stability Analysis} \label{sec:LSA}
We first perform a global linear stability analysis (LSA) to inspect the stability properties and long time-horizon dynamics of the flow. We employ a modal ansatz harmonic in $\theta$ and $t$ for the fluctuating quantities:
\begin{equation}\label{eq:ansatz}
    \bfq'(x,r,\theta,t) = \hbfq(x,r) \, \mathrm{e}^{i(m\theta - \omega t)} + \textrm{c.c.}, 
\end{equation}
where $\hbfq$ is the complex spatial mode shape, $m$ is the azimuthal wavenumber, $\omega = \omega_r + i \omega_i$ is the complex frequency and \enquote{$\textrm{c.c.}$} denotes the complex conjugate.

In the absence of forcing ($\bff'=0$), the perturbation equations (\ref{eq:NS_perturbation1}--\ref{eq:NS_perturbation2}) with appropriate boundary conditions reduce to the generalized eigenvalue problem:
\begin{equation}\label{eq:GEVP}
-i\omega\, \boldsymbol{\mathbf{B}}
\left(\begin{array}{@{}c@{}}\hbfu\\ 
\hat{p}\end{array}\right) = \boldsymbol{\mathbf{L}}\,\left(\begin{array}{@{}c@{}}\hbfu\\ 
\hat{p}\end{array}\right),
\end{equation}
where $\boldsymbol{\mathbf{B}}$ is the mass matrix arising from the spatial discretization, and $\boldsymbol{\mathbf{L}}$ is the 2D linearized Navier–Stokes operator. 
To prevent spurious reflections at the domain inlet and outlet, we apply a quadratic sponge region to damp all fluctuations.
We solve this system using the finite-element linear solver FELiCS \citep{kaiser_felics_2023} over a 2D grid of $8.2\times10^4$ triangular elements. The eigenvalues and eigenvectors are computed with the SLEPc library
, while the Arnoldi \textit{shift-and-invert} algorithm returns the n-th closest eigenvalues to a prescribed initial guess.

\begin{figure}[t]
\centering
\includegraphics[width=1\linewidth]{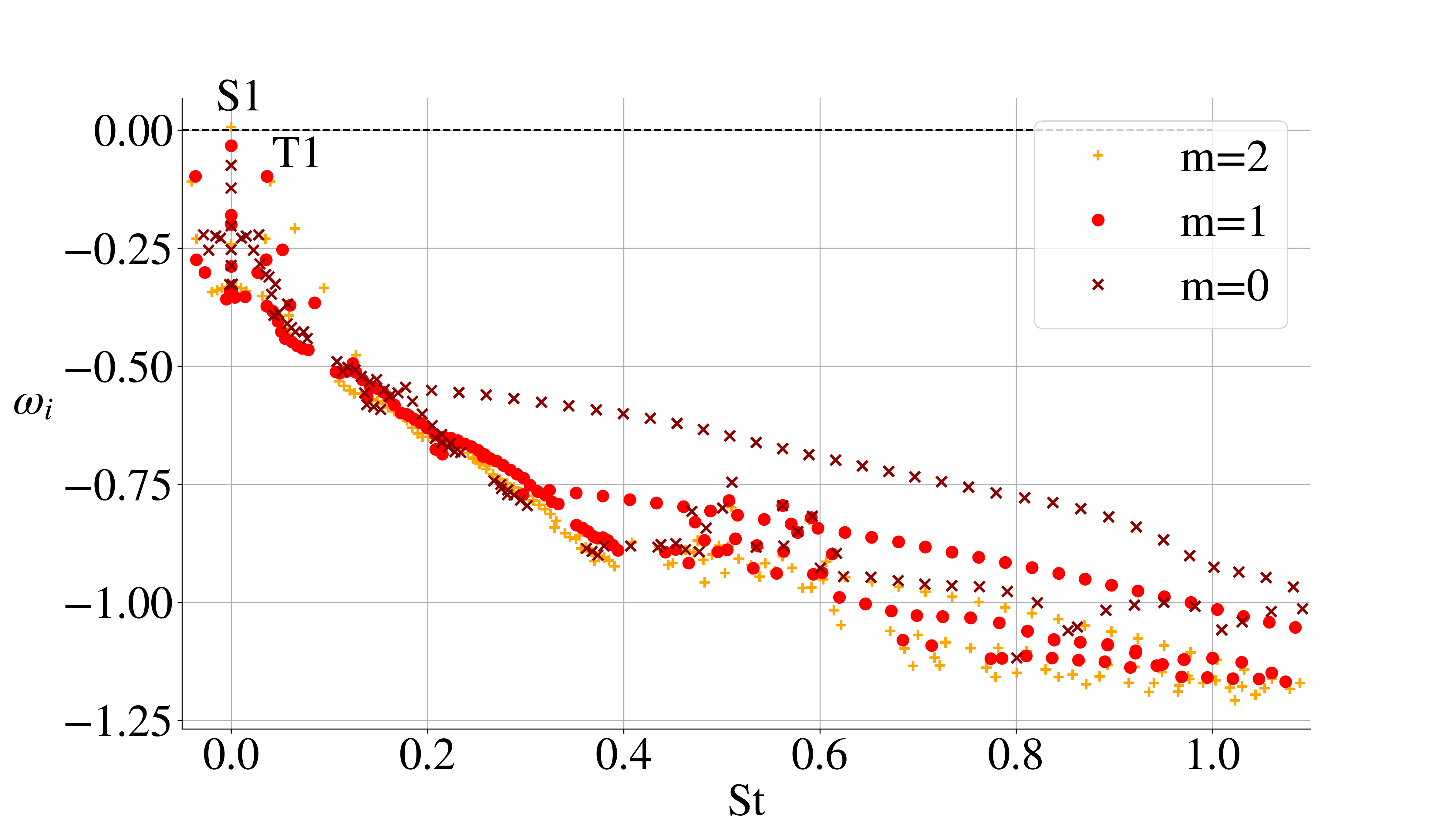}
\caption{Eigenvalue spectrum of the global linear stability analysis for the first three azimuthal modes. The first stationary and traveling modes are marked S1 and T1, respectively.}
\label{fig:LSA_spectrum}
\end{figure}

\begin{figure}%
    \centering
    \def\svgwidth{1\linewidth}
\begingroup%
  \makeatletter%
  \providecommand\color[2][]{%
    \errmessage{(Inkscape) Color is used for the text in Inkscape, but the package 'color.sty' is not loaded}%
    \renewcommand\color[2][]{}%
  }%
  \providecommand\transparent[1]{%
    \errmessage{(Inkscape) Transparency is used (non-zero) for the text in Inkscape, but the package 'transparent.sty' is not loaded}%
    \renewcommand\transparent[1]{}%
  }%
  \providecommand\rotatebox[2]{#2}%
  \newcommand*\fsize{\dimexpr\f@size pt\relax}%
  \newcommand*\lineheight[1]{\fontsize{\fsize}{#1\fsize}\selectfont}%
  \ifx\svgwidth\undefined%
    \setlength{\unitlength}{536.94804995bp}%
    \ifx\svgscale\undefined%
      \relax%
    \else%
      \setlength{\unitlength}{\unitlength * \real{\svgscale}}%
    \fi%
  \else%
    \setlength{\unitlength}{\svgwidth}%
  \fi%
  \global\let\svgwidth\undefined%
  \global\let\svgscale\undefined%
  \makeatother%
  \begin{picture}(1,0.10897266)%
    \lineheight{1}%
    \setlength\tabcolsep{0pt}%
    \put(0,0){\includegraphics[width=\unitlength,page=1]{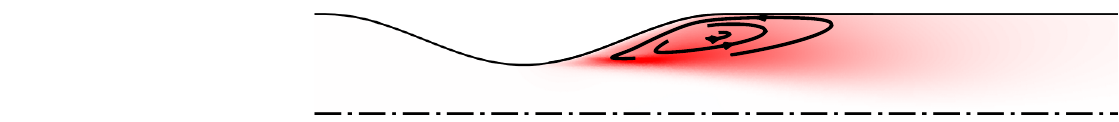}}%
    \put(-0.00245568,0.03813938){\color[rgb]{0,0,0}\makebox(0,0)[lt]{\lineheight{0}\smash{\begin{tabular}[t]{l}Real($\hat{u}_x^{S1}$)\end{tabular}}}}%
  \end{picture}%
\endgroup%

    \caption{Axial velocity component of the stationary mode S1 at $m=2, \mathrm{St} = 0$. The color map is saturated at maximum absolute value of the magnitude.}
    \label{fig:LSA_mode_shape} 
\end{figure}

Figure~\ref{fig:LSA_spectrum} shows the LSA spectrum for $m=0,1,2$  as a function of the Strouhal number $\mathrm{St} = f d_t / U_t$.
At intermediate frequencies $\mathrm{St} > 0.2$, the spectrum is composed of stable continuous branches.
Along the zero-frequency axis ($\mathrm{St} = 0$), a discrete stationary mode lies near the neutral threshold for all considered values of $m \geq 1$. This stationary eigenmode is most amplified for $m=2$ and has a positive growth rate for $m=2$ and $m=3$ (not shown here) only. All other $m$ (including $m\geq 4$, omitted in figure~\ref{fig:LSA_spectrum} for clarity) exhibit a damped stationary mode. We call S1 this least stable stationary mode for every $m \geq 1$.
The axial velocity component of the leading stationary mode for $m=2$ is shown in figure~\ref{fig:LSA_mode_shape}, while the other two components with magnitude orders close to zero are omitted for clarity. It is primarily localized in the shear-layer and inside the separation bubble. Note that the stationary modes S1 for $1 \leq m \leq 6$ exhibit similar mode shapes, whereas the $m=0$ stationary modes yield velocity components approaching machine precision. 

In their LSA of the same flow at $Re = 750$, \citet{sherwin2005three} identified stationary eigenmodes for $m\geq1$ with similar structure, finding $m=1$ to be unstable while $m\geq2$ modes remained stable. They described it as a mild Coanda-type instability, in which the jet attaches to one side of the wall.
However, for this turbulent flow, both the $m=2$ and the $m=3$ modes amplify, meaning that the instability is a combination of these two types of jet attachment downstream of the recirculation. 
Such a dominant $m=2$ stationary eigenmode has been identified as the leading unstable mode in LSA of stenosis with area reduction below $60\%$ \citep{griffith2008steady}, and is described as a pinching of the jet.

Since then, similar stationary eigenmodes have been documented in several separated flows with recirculation bubbles, including pressure-induced separation bubbles \citep{wu2020spatio, cura2024low, fuchs2025standing} and separated flows over a Gaussian bump \citep{klopsch2025enabling}.
Recent investigations have associated this stationary mode with a centrifugal instability \citep{fuchs2025standing}.
While these findings are restricted to non-axisymmetric configurations, the shared characteristics with our S1 mode lead us to hypothesize that the same fundamental mechanism is at work in the present axisymmetric configuration.

The maximum growth rate of the S1 mode across wavenumbers is primarily governed by the magnitude of the reversed flow within the separation bubble. Previous studies have shown that it becomes unstable for peak reversed-flow velocities above $1.1\%$ of the freestream velocity in turbulent separation bubbles \citep{cura2025linear}. The present flow exhibits a peak reversed-flow velocity of $0.17 \,U_t$, which is consistent with the observed amplification of S1 for some azimuthal wavenumbers.

In summary, two descriptions of this stationary instability describe the same instability. It could be view either as the jet reattaching downstream of the recirculation bubble (Coanda-type global instability mode), or as a deformation of the recirculation bubble. 

As \citet{griffith2008steady} demonstrated, convective instability in the shear-layer drives the primary route to turbulence in this flow. Thus, the forced dynamics of the system needs to be investigated to complement the LSA and fully characterize our turbulent case.

\subsection*{Resolvent Analysis}
We use resolvent analysis to identify non-modal amplification mechanisms, characterizing the input-output response of the linearized system to harmonic forcing. 

Transforming the perturbation equations (\ref{eq:NS_perturbation1}--\ref{eq:NS_perturbation2}) into the frequency domain using the modal ansatz (\ref{eq:ansatz}) yields the resolvent operator:
\begin{equation}
\hbfu = \mathcal{R}\, \hbff, \quad\text{where} \quad \mathcal{R} = \mathbf{D}_r\mathbf{P}^T \,( -i\omega \boldsymbol{\mathbf{B}} - \boldsymbol{\mathbf{L}} )^{-1} \mathbf{P}\mathbf{D}_f.
\end{equation}
Here, $\mathbf{P}$ removes the forcing in the mass equation and the pressure component. The projection operators $\mathbf{D}_f$ and $\mathbf{D}_r$ restrict the input and output spatial regions to the immediate post-stenotic zone $x/d_t \in [-2, 4]$ \citep{towne_spectral_2018}, ensuring that the sponge region does not interfere with the computed resolvent modes.

A singular value decomposition of $\mathcal{R}$ at a given $(m, \omega)$ yields the optimal forcing mode $\hbff^{(j)}(m, \omega)$ and response mode $\hbfu^{(j)}(m, \omega)$ as the right and left singular vectors, respectively. The corresponding singular values $\sigma^{(j)}(m, \omega)$ quantify the amplification of the $j^{\mathrm{th}}$ resolvent mode.
The singular value decomposition is solved using the same mean field, numerical framework, mesh, and boundary conditions as the LSA.

\begin{figure}[t]
\centering
\includegraphics[width=1\linewidth]{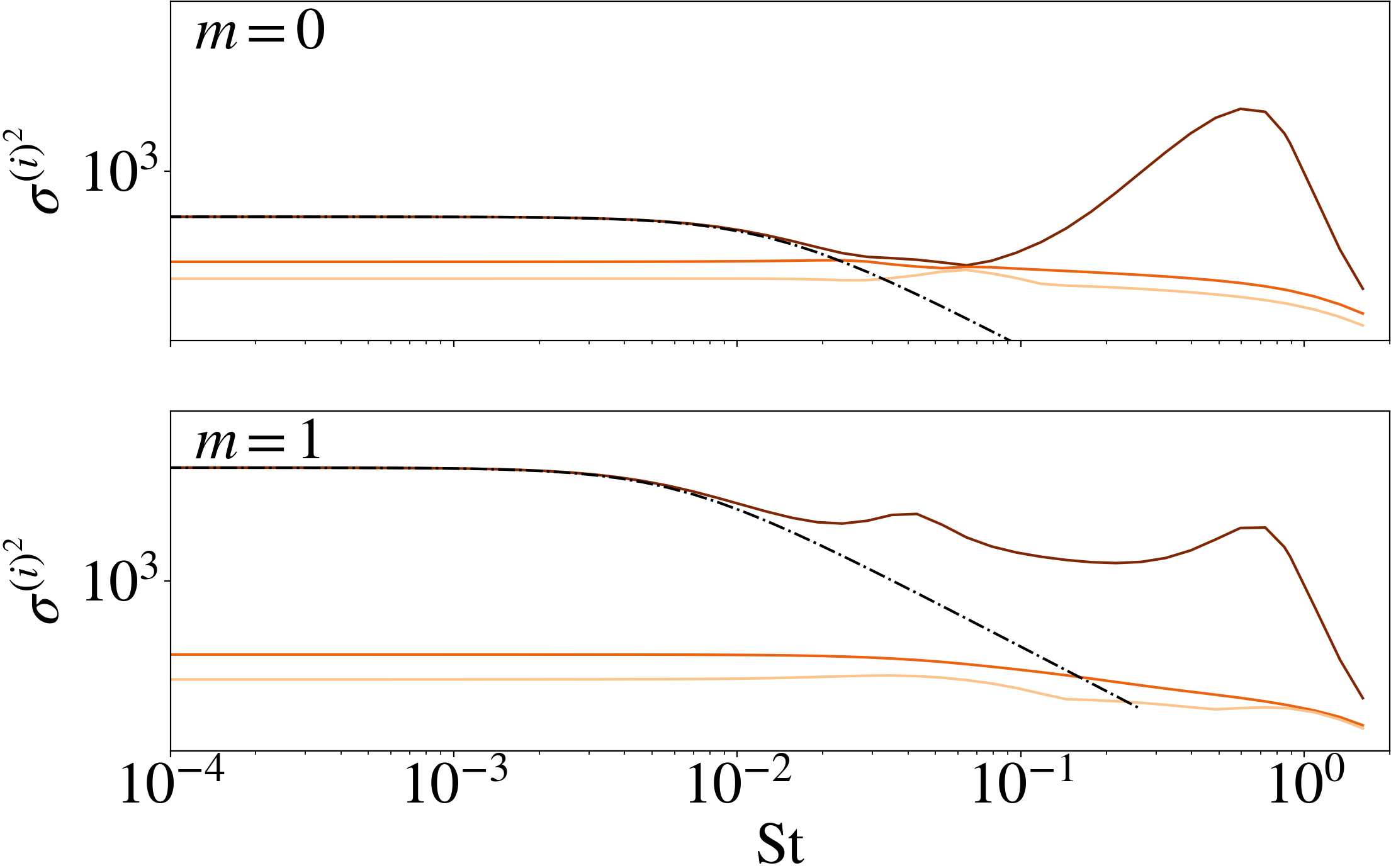}
\caption{Resolvent gain spectra for the first four resolvent modes for the first two azimuthal wavenumbers. The low-pass filter model from \citet{bugeat2022low} is added in dashed lines.}
\label{fig:RA_spectrum}
\vspace{-0.4cm}
\end{figure}

As the linear operator is stable for $m = 0$ and $m = 1$, the resolvent operator is well-defined for these azimuthal modes.
Figure~\ref{fig:RA_spectrum} shows the squared gains of the first four resolvent modes as a function of the Strouhal number for $m=0,1$.
Two main amplification regions with high gain separation appear.
In the low-frequency regime ($\mathrm{St} < 0.01$), the gains reveal a low-pass filter behavior with an apparent cutoff frequency, which is known to emerge from the identified S1 eigenmode \citep{robinet2007bifurcations, cura2024low}.
The influence of S1 on the resolvent gains has been modeled by \citet{bugeat2022low}. The low-frequency amplification caused by the stationary eigenvalue at first order is:
\begin{equation} 
    \sigma (\mathrm{St}) = \frac{\sigma^{(0)}(St\rightarrow 0)}{\sqrt{1+ \left( \dfrac{\mathrm{St}}{\omega_i^{(S1)}/2 {\rm \pi}} \right)^2}}, 
\end{equation}
where $\sigma^{(0)}(St\rightarrow 0)$ is the value of the optimal gain when $\mathrm{St}$ approaches zero, and $\omega_i^{(S1)}$ is the growth rate of the S1 mode.
The modal amplification triggered by the S1 mode only is plotted in dashed lines in figure~\ref{fig:RA_spectrum} and closely matches the low-frequency gains. This confirms that the low-frequency receptivity is associated with the forced modal response of S1.

The small amplification peak at $m=1, \, \mathrm{St}=0.04$ in figure~\ref{fig:RA_spectrum}(b)  is also modal and corresponds to the resonance of the most weakly damped traveling mode T1 [see figure~\ref{fig:LSA_spectrum}] at the same frequency.

In the high-frequency regime, a broadband amplification dominates the dynamics. It is centered around $\mathrm{St} = 0.59$ for $m=0$ and $\mathrm{St} = 0.73$ for $m=1$. As no eigenvalue lies near the neutral threshold in the LSA spectrum for this frequency, this amplification is a pseudo-resonance i.e. it arises from the linear superposition of globally stable modes.
As opposed to the transient growth analysis results at $Re = 750$ from \citet{blackburn2008convective}, the $m=0$ perturbations, rather than the $m=1$ modes, exhibit the greatest amplification. This 2D characteristic corresponds to the axisymmetric rollers that develop within the shear-layer at $\mathrm{St}=0.5$, which experiments \citep{vetel2008asymmetry} and direct numerical simulations \citep{varghese2007direct} observed at $Re=1000$.
%

\begin{figure}%
    \centering
    \def\svgwidth{1\linewidth}
    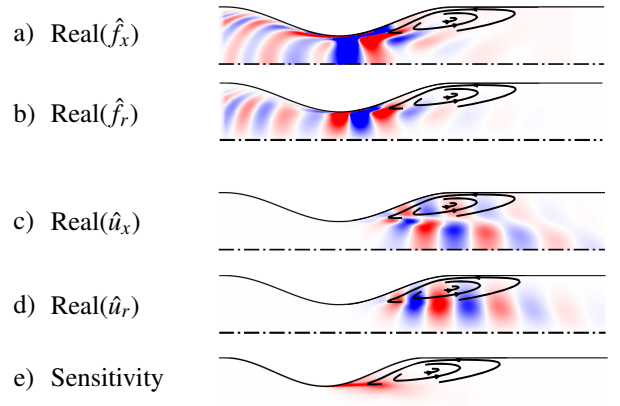
    \caption{Mode shapes of the optimal resolvent forcing (a,b) and response (c,d) at $m=0, \, \mathrm{St}=0.59$. The panels show the real part of the axial and radial velocity components, together with the recirculation bubble streamlines. The colormap is saturated at the maximum absolute amplitude for each mode and amplified by a factor of $100$ for the forcing. Panel (e) illustrates the structural sensitivity of this resolvent mode.}
    \label{fig:Mode_shapes_RA}
\end{figure}

The forcing and response optimal mode shapes at $m=0, \,\mathrm{St}=0.59$ are plotted in figure~\ref{fig:Mode_shapes_RA}(a-d).
The optimal forcing is located upstream of the recirculation. Its orientation and location is similar to the optimal initial perturbation computed in the transient growth analysis of the same mean flow \citet{griffith2010convective}. The concentration of the forcing at the separation point highlights the strong influence of this region of the flow on the amplified dynamics. It leads to a downstream response that amplifies in the separated shear-layer and reaches a maximum amplitude at the recirculation center point where backflow is the strongest. The mode shape and frequency are reminiscent of the convective Kelvin–Helmholtz instability in turbulent jets at similar ($m, \, \mathrm{St}$) \citep{pickering2020lift}.
This prevalent non-normal amplification is typical in amplifier flows and reflects the convective instability of the separated shear-layer \citep{garnaud2013preferred}. 

The overlap region between the forcing and response mode reveal the region where small perturbations of the flow have the largest effect on the resolvent gain. The resolvent-base structural sensitivity of a mode is defined in \citet{pickering2020lift} by taking the Frobenius norm of the sensitivity tensor 
$\boldsymbol{S}_{ij} = \sigma ^2 \text{Real}(\hbff_i\hbfu_j^*)$. The resulting sensitivity is shown in figure~\ref{fig:Mode_shapes_RA}(e). It is localized in the shear-layer close to the separation point. In order to dampen this dominant amplification, the most effective way would be to act on this sensitive separation flow region. This implies that this convective amplification is sensitive to the stenosis degree, as highlighted in \citet{griffith2008steady}.

\section*{CONCLUSION}
Ultimately, the presented methodology benefits from the non-invasive 4D-flow MRI velocity measurements and enhance the data to infer driving instabilities from time-average velocity measurements alone. 
To overcome the biases and noise in the standard 4D-flow MRI measurements, a two-step PINN approach corrects the displacement artifacts and assimilates the unknown mean pressure and eddy viscosity fields. This process yields a RANS-consistent base state suitable for stability analysis.
The global LSA of the assimilated mean flow identified a stationary eigenmode for all $m\geq 1$, with a positive growth rate for $m=2$ and $m=3$. This eigenmode influences the forced response of the system in a range of frequencies dictated by its growth rate. 
For $\mathrm{St} \in [0.1, 1]$, resolvent analysis identifies a broadband pseudo-resonance associated with the Kelvin-Helmholtz instability of the separated shear-layer, with maximal amplification for $m=0$.

\bibliographystyle{tsfp}

\begin{thebibliography}{29}
\expandafter\ifx\csname natexlab\endcsname\relax\def\natexlab#1{#1}\fi

\bibitem[Arzani {\em et~al.\/}(2021)]{arzani_uncovering_2021}
Arzani, A. {\em et~al.\/} 2021 Uncovering near-wall blood flow from sparse data with physics-informed neural networks. {\em Physics of Fluids\/} {\bf 33}~(7), 071905.

\bibitem[Blackburn {\em et~al.\/}(2008)]{blackburn2008convective}
Blackburn, H.~M. {\em et~al.\/} 2008 Convective instability and transient growth in flow over a backward-facing step. {\em Journal of Fluid Mechanics\/} {\bf 603}, 271--304.

\bibitem[Bugeat {\em et~al.\/}(2022)]{bugeat2022low}
Bugeat, B. {\em et~al.\/} 2022 Low-frequency resolvent analysis of the laminar oblique shock wave/boundary layer interaction. {\em Journal of Fluid Mechanics\/} {\bf 942}, A43.

\bibitem[Busch {\em et~al.\/}(2013)]{Busch2013}
Busch, J. {\em et~al.\/} 2013 {Reconstruction of divergence-free velocity fields from cine 3D phase-contrast flow measurements}. {\em Magnetic Resonance in Medicine\/} {\bf 69}~(1), 200--210.

\bibitem[Cura {\em et~al.\/}(2024)]{cura2024low}
Cura, C. {\em et~al.\/} 2024 On the low-frequency dynamics of turbulent separation bubbles. {\em Journal of Fluid Mechanics\/} {\bf 991}, A11.

\bibitem[Cura {\em et~al.\/}(2025)]{cura2025linear}
Cura, C. {\em et~al.\/} 2025 Linear modeling of a family of turbulent separation bubbles. {\em Physical Review Fluids\/} {\bf 10}~(11), 114607.

\bibitem[Dillinger {\em et~al.\/}(2020)]{dillinger2020limitations}
Dillinger, H. {\em et~al.\/} 2020 On the limitations of echo planar 4d flow mri. {\em Magnetic resonance in medicine\/} {\bf 84}~(4), 1806--1816.

\bibitem[Fuchs {\em et~al.\/}(2025)]{fuchs2025standing}
Fuchs, M.~L. {\em et~al.\/} 2025 A standing-wave model for the low-frequency dynamics of a turbulent separation bubble.

\bibitem[Garnaud {\em et~al.\/}(2013)]{garnaud2013preferred}
Garnaud, X. {\em et~al.\/} 2013 The preferred mode of incompressible jets: linear frequency response analysis. {\em Journal of Fluid Mechanics\/} {\bf 716}, 189--202.

\bibitem[Griffith {\em et~al.\/}(2008)]{griffith2008steady}
Griffith, M.~D. {\em et~al.\/} 2008 Steady inlet flow in stenotic geometries: convective and absolute instabilities. {\em Journal of fluid mechanics\/} {\bf 616}, 111--133.

\bibitem[Griffith {\em et~al.\/}(2010)]{griffith2010convective}
Griffith, M.~D. {\em et~al.\/} 2010 Convective instability in steady stenotic flow: optimal transient growth and experimental observation. {\em Journal of fluid mechanics\/} {\bf 655}, 504--514.

\bibitem[Kaiser {\em et~al.\/}(2023)]{kaiser_felics_2023}
Kaiser, T.L. {\em et~al.\/} 2023 {FELiCS}: {A} {Versatile} {Linearized} {Solver} {Addressing} {Dynamics} in {Multi}-{Physics} {Flows}. In {\em {AIAA} {AVIATION} 2023 {Forum}\/}. San Diego, CA and Online: American Institute of Aeronautics and Astronautics.

\bibitem[Klopsch {\em et~al.\/}(2025)]{klopsch2025enabling}
Klopsch, R. {\em et~al.\/} 2025 Enabling resolvent analysis through assimilation of experimental mean flows with physics-informed neural networks: A case study on the boeing gaussian bump. In {\em AIAA AVIATION FORUM AND ASCEND 2025\/}, p. 3604.

\bibitem[Ku(1997)]{ku1997blood}
Ku, David~N 1997 Blood flow in arteries. {\em Annual review of fluid mechanics\/} {\bf 29}~(1), 399--434.

\bibitem[Markl {\em et~al.\/}(2012)]{markl20124d}
Markl, M. {\em et~al.\/} 2012 4d flow mri. {\em Journal of Magnetic Resonance Imaging\/} {\bf 36}~(5), 1015--1036.

\bibitem[McKeon {\em et~al.\/}(2010)]{mckeon2010critical}
McKeon, B.~J. {\em et~al.\/} 2010 A critical-layer framework for turbulent pipe flow. {\em Journal of Fluid Mechanics\/} {\bf 658}, 336--382.

\bibitem[Pickering {\em et~al.\/}(2020)]{pickering2020lift}
Pickering, E. {\em et~al.\/} 2020 Lift-up, kelvin--helmholtz and orr mechanisms in turbulent jets. {\em Journal of Fluid Mechanics\/} {\bf 896}, A2.

\bibitem[Raissi {\em et~al.\/}(2019)]{raissi2019physics}
Raissi, M. {\em et~al.\/} 2019 Physics-informed neural networks: A deep learning framework for solving forward and inverse problems involving nonlinear partial differential equations. {\em Journal of Computational physics\/} {\bf 378}, 686--707.

\bibitem[Robinet(2007)]{robinet2007bifurcations}
Robinet, J-Ch 2007 Bifurcations in shock-wave/laminar-boundary-layer interaction: global instability approach. {\em Journal of Fluid Mechanics\/} {\bf 579}, 85--112.

\bibitem[von Saldern {\em et~al.\/}(2022)]{von_saldern_mean_2022}
von Saldern, J.G.R. {\em et~al.\/} 2022 Mean flow data assimilation based on physics-informed neural networks. {\em Physics of Fluids\/} {\bf 34}~(11), 115129.

\bibitem[Schmid {\em et~al.\/}(2001)]{schmid2001stability}
Schmid, P.~J. {\em et~al.\/} 2001 {\em Stability and transition in shear flows\/}, , vol. 142. Springer Science \& Business Media.

\bibitem[Sherwin {\em et~al.\/}(2005)]{sherwin2005three}
Sherwin, S. {\em et~al.\/} 2005 Three-dimensional instabilities and transition of steady and pulsatile axisymmetric stenotic flows. {\em Journal of Fluid Mechanics\/} {\bf 533}, 297--327.

\bibitem[Thunberg {\em et~al.\/}(2002)]{Thunberg2002}
Thunberg, P. {\em et~al.\/} 2002 {Correction for displacement artifacts in 3D phase contrast imaging}. {\em Journal of Magnetic Resonance Imaging\/} {\bf 16}~(5), 591--597.

\bibitem[Towne {\em et~al.\/}(2018)]{towne_spectral_2018}
Towne, A {\em et~al.\/} 2018 Spectral proper orthogonal decomposition and its relationship to dynamic mode decomposition and resolvent analysis. {\em Journal of Fluid Mechanics\/} {\bf 847}, 821--867.

\bibitem[Varghese {\em et~al.\/}(2007)]{varghese2007direct}
Varghese, S. {\em et~al.\/} 2007 Direct numerical simulation of stenotic flows. part 1. steady flow. {\em Journal of Fluid Mechanics\/} {\bf 582}, 253--280.

\bibitem[V{\'e}tel {\em et~al.\/}(2008)]{vetel2008asymmetry}
V{\'e}tel, J. {\em et~al.\/} 2008 Asymmetry and transition to turbulence in a smooth axisymmetric constriction. {\em Journal of Fluid Mechanics\/} {\bf 607}, 351--386.

\bibitem[Villi{\'e} {\em et~al.\/}(2025)]{villie2025physics}
Villi{\'e}, A. {\em et~al.\/} 2025 Physics-informed neural networks for enhancing medical flow magnetic resonance imaging: Artifact correction and mean pressure and reynolds stresses assimilation. {\em Physics of Fluids\/} {\bf 37}~(2).

\bibitem[Von~Saldern {\em et~al.\/}(2024)]{vonsaldernRoleEddyViscosity2024}
Von~Saldern, J.G.R. {\em et~al.\/} 2024 On the role of eddy viscosity in resolvent analysis of turbulent jets. {\em Journal of Fluid Mechanics\/} {\bf 1000}, A51.

\bibitem[Wu {\em et~al.\/}(2020)]{wu2020spatio}
Wu, W. {\em et~al.\/} 2020 Spatio-temporal dynamics of turbulent separation bubbles. {\em Journal of Fluid Mechanics\/} {\bf 883}, A45.

\end{thebibliography}

\providecommand{\noopsort}[1]{}\providecommand{\singleletter}[1]{#1}%

%
%
%
%
%


\end{document}